\documentclass[11pt]{article}

\usepackage{amsmath,amssymb,graphicx}

\renewcommand{\title}[1]{%
	\begin{center} \Large \bf #1 \end{center}%
	}
\renewcommand{\author}[2]{%
	\begin{center} { #1}  \vspace{2mm}\\ %
	  \it #2%
	\end{center}%
	\addvspace{\baselineskip}%
	}

\begin{document}
\newpage
\setcounter{section}{0}
\setcounter{equation}{0}
\setcounter{figure}{0}
\baselineskip 5mm
\title{%
Discrimination of Models Including Doubly Charged Scalar Bosons
by Using Tau Lepton Decay Distributions%
\footnote{
 This talk
in the CST-MISC Joint International Symposium on Particle Physics
is based on a work in Ref.~\cite{Sugiyama:2012yw}
in collaboration with Koji Tsumura and Hiroshi Yokoya.
}
}%
\author{%
Hiroaki Sugiyama
}{%
Maskawa Institute for Science and Culture,
Kyoto Sangyo University, Kyoto 603-8555, Japan
}%
\section{Introduction}

 Measurements of neutrino oscillations
have shown the existence of non-zero masses of neutrinos
which are regarded as massless particles
in the standard model of particle physics~(SM).
 The SM must be extended such that
the non-zero neutrino masses are accommodated.
 Some of such extended models for the neutrino mass
involve the doubly-charged scalar boson~$H^{--}$.

 In the Higgs triplet model~(HTM)~\cite{Ref:HTM}, for example,
an $\text{SU}(2)_L$-triplet scalar field $\Delta$
with the hypercharge $Y=1$
is introduced to the SM
in order to accommodate non-zero neutrino masses.
 The $\Delta$ can be expressed
in an adjoint representation as
\begin{eqnarray}
\Delta
\equiv
 \begin{pmatrix}
  \Delta^+/\sqrt{2} & \Delta^{++}\\
  \Delta^0 & -\Delta^+/\sqrt{2}
 \end{pmatrix} .
\end{eqnarray}
 The scalar field $\Delta$ has
the following Yukawa interaction with
the $\text{SU}(2)_L$-doublet field
$L_\ell \equiv (\nu_{\ell L}^{} , \ell_L^{})^T$
of left-handed leptons in the flavor basis~($\ell = e, \mu, \tau$):
\begin{eqnarray}
{\mathcal L}_{\text{Yukawa}}^{\text{HTM}}
=
 - (h_M^{})_{\ell\ell^\prime}\,
 \overline{L_\ell^c}\, \epsilon\, \Delta\, L_{\ell^\prime} + \text{h.c.},
\label{Eq:HTM-Yukawa}
\end{eqnarray}
where $h_M^{}$ is a symmetric matrix of Yukawa coupling constants,
and $\epsilon$ denotes the completely antisymmetric tensor
for the $\text{SU}(2)_L$ indices.
 The Majorana neutrino mass matrix $M_\nu^\text{HTM}$ in the flavor basis
is simply given by $M_\nu^{\text{HTM}} = 2 h_M^{} \langle \Delta^0 \rangle$,
where $\langle \Delta^0 \rangle$ is
the vacuum expectation value of $\Delta^0$.
 It is clear that
the doubly-charged scalar $\Delta^{--}$
has an interaction $\Delta^{--} \overline{\ell_L^{}} (\ell_L^\prime)^c$,
and the leptonic decay is $\Delta^{--} \to \ell_L^{} \ell_L^\prime$.

 The Zee-Babu model~(ZBM)~\cite{Ref:ZBM}
is another simple example of models to generate neutrino masses,
where doubly-charged scalar boson is introduced.
 In the ZBM,
the SM is extended by introducing
two $\text{SU}(2)_L$-singlet scalar bosons
$s^{++}$ and $s^+$
whose hypercharges are $Y=2$ and $1$, respectively.
 Yukawa interactions of these new scalar bosons with leptons
are given by
\begin{eqnarray}
{\mathcal L}_{\text{Yukawa}}^{\text{ZBM}}
=
 - (Y_s)_{\ell\ell^\prime}\,
   \overline{(\ell_R)^c}\, \ell^\prime_R\, s^{++}
 - (Y_a)_{\ell\ell^\prime}\,
   \overline{L_\ell^c}\, \epsilon\, L_{\ell^\prime}\, s^+
 + \text{h.c.} ,
\end{eqnarray}
where $Y_s$~($Y_a$) is a symmetric (an antisymmetric) matrix
of Yukawa coupling constants.
 Tiny Majorana neutrino masses are
generated at the two-loop level
where charged leptons and new scalar bosons
are involved in the loop.
 In contrast with the HTM,
the leptonic decay of the doubly-charged scalar boson
is $s^{--} \to \ell_R^{} \ell_R^\prime$.

 In general,
there are two possibilities for
the Yukawa interaction of $H^{--}$ with charged leptons,
$H^{--}_X \overline{\ell_X^{}} (\ell_X^\prime)^c$~($X=L, R$),
where $H^{--}_L$~($H^{--}_R$) stands for
$H^{--}$ whose Yukawa interaction is only with
left-handed~(right-handed) charged leptons.
 A simple example of $H^{--}_L$ is $\Delta^{--}$ in the HTM
while $s^{--}$ in the ZBM
is a simple example of $H^{--}_R$.
 It will be important to distinguish between $H^{--}_L$ and $H^{--}_R$,
namely between $\ell_L^{}$ and $\ell_R^{}$ produced by the $H^{--}$ decay,
in order to discriminate models for generating neutrino masses
when a $H^{--}$ is discovered in the future.

 The tau lepton decays within a detector in a collider experiment
while the muon is too light to decay within the detector.
 It is known that
decay products have information on the polarization
of the parents tau lepton%
~\cite{Ref:tau-pol, Ref:tau-pol-Higgs, Ref:tau-pol-SUSY}.
 Since the tau lepton is heavier than the pion,
it has a simple two-body decay channel
which is especially useful for the extraction
of information on the polarization of the tau lepton.
 The angular distribution of the daughter pions
in the rest flame of the tau lepton
is translated into the energy distribution of the pion
in the collinear limit, $E_\tau \gg m_\tau^{}$,
where $E_\tau$ and $m_\tau^{}$ are
the energy and the mass of the tau lepton, respectively.
 In the collinear limit,
energy distributions $F_X^\pi(z)$
for $\pi^-$ produced via
the two-body decay of the polarized tau lepton $\tau_X^{}$~($X=L, R$)
are known well as
\begin{eqnarray}
 F_L^\pi(z) = 2 (1-z) , \quad
%
%
 F_R^\pi(z) = 2z,
\end{eqnarray}
where $z \equiv E_\pi/E_\tau$ for the pion energy $E_\pi$.
 As it is naively expected,
$\tau_L^{} \to \nu_L^{} \pi^-$
tends to produce a soft pion
while a pion produced via $\tau_R^{} \to \overline{\nu_L^{}} \pi^-$
tends to be a hard one.
 Note that
the collinear limit is always reliable
when the tau lepton is produced by the decay of $H^{--}$
because $H^{--}$ is much heavier than the tau lepton.
 In this talk,
we see that
$H^{--}_L$ and $H^{--}_R$ can be distinguished
by measuring energy distributions of daughter charged pions
of tau leptons produced by decays of $H^{--}$.

\section{Results}

 Hereafter,
$\ell$ denotes $e$ and $\mu$,
and $\ell\ell$ means not only $ee$ and $\mu\mu$ but also $e\mu$.
 We deal only with the pair production of $H^{--}$.

\begin{figure}[t]
\begin{center}
\includegraphics[origin=c, angle=0, scale=0.58]{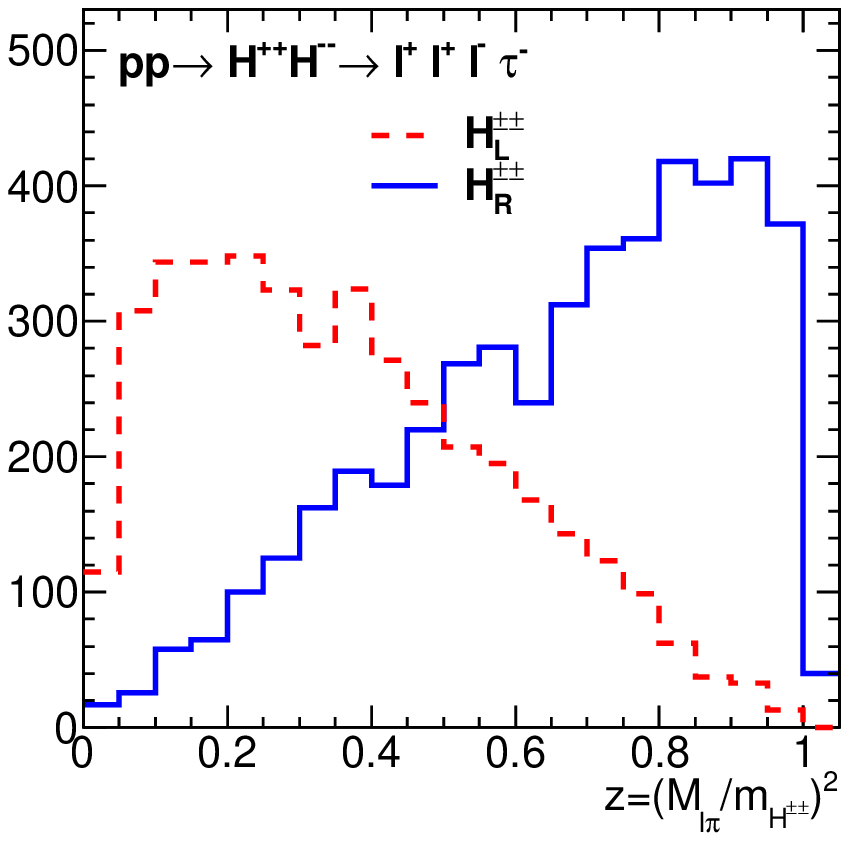}
\hspace*{1mm}
\includegraphics[origin=c, angle=0, scale=0.58]{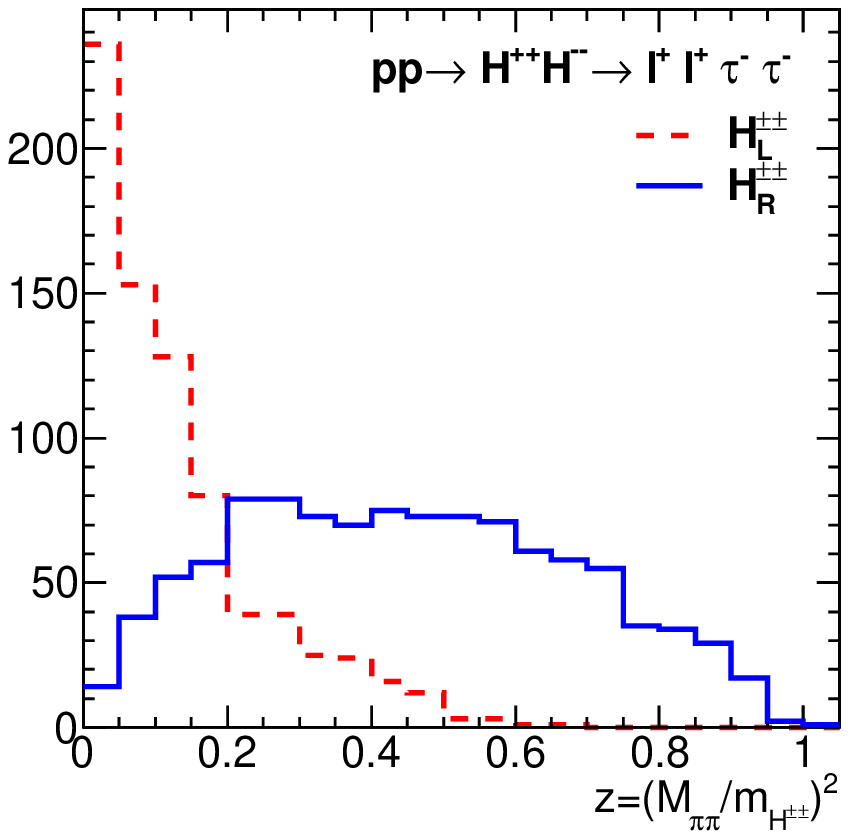}
\hspace*{1mm}
\includegraphics[origin=c, angle=0, scale=0.86]{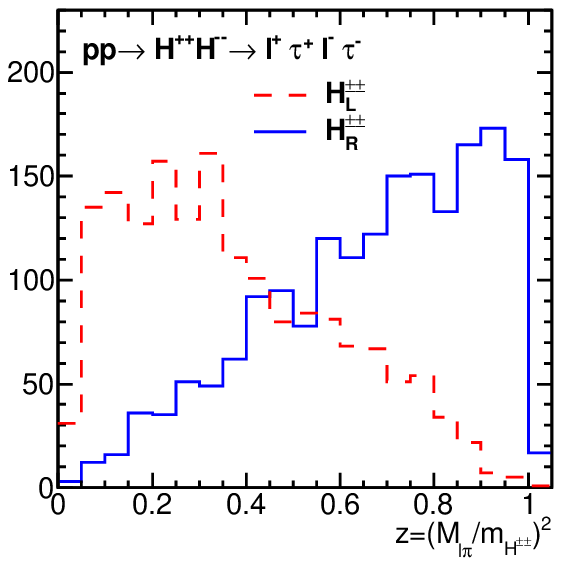}
\caption{
 Distributions of charged pions in our simulation.
 These figures are taken from Ref.~\cite{Sugiyama:2012yw}.
}
\label{Fig:figures}
\end{center}
\end{figure}

 Let us consider first the case
where $H^{--}$ decays into $\ell\ell$ and $\ell\tau$
with sizable branching ratios.
 The $H^{--}$ can be discovered by using
a pair of same-signed $\ell$
whose invariant mass $M_{\ell\ell}$ has a peak
at the mass $m_{H^{\pm\pm}}^{}$ of $H^{--}$.
 Then,
$pp \to H^{++} H^{--}
\to \overline{\ell}\, \overline{\ell}\, \ell\, \tau
\to \overline{\ell}\, \overline{\ell}\, \ell\, \pi^-\, \nu$
is the most useful mode to determine the tau polarization.
 The background for the signal will be significantly removed
by requiring the existence of a pair of same-signed charged leptons
with $M_{\ell\ell} \simeq m_{H^{\pm\pm}}^{}$.
 We see that the pion energy fraction $z$~($\equiv E_\pi/E_\tau$)
can be calculated as $z = M_{\pi\ell}^2/m_{H^{\pm\pm}}^2$
in the collinear limit,
where $M_{\pi\ell}$ is the invariant mass
of a same-singed charged pair of a pion and a charged lepton.
 The distribution of charged pions
with respect to $z$~($= M_{\pi\ell}^2/m_{H^{\pm\pm}}^2$)
in our simulation
is shown in Fig.~\ref{Fig:figures}~(left).
 There remains a remarkable difference
between distributions
for $H^{--}_L$~(red dashed line) and $H^{--}_R$~(blue solid line)
even after the event selection in our simulation.
 In order to distinguish between $H^{--}_L$ and $H^{--}_R$,
it is sufficient to observe the difference
between numbers of events for $z < 0.5$ and $0.5 < z$.
 The difference of these numbers is about $2000$
in Fig.~\ref{Fig:figures}~(left)
where we generated $3\times 10^4$ events
of $pp \to H^{++} H^{--}
\to \overline{\ell}\, \overline{\ell}\, \ell\, \tau$
followed only by the hadronic $\tau$ decays.%
\footnote
{
 The actual branching ratio of the hadronic $\tau$ decays is $64.8\,\%$,
and the branching ratio for $\tau \to \pi^- \nu$ is $10.8\,\%$.
 Thus, $17\,\%$ of the tau lepton in our simulation
decays into $\pi^-\nu$.
}
 The difference will be $O(10)$ events
even if the LHC produces only $O(100)$ events of
$pp \to H^{++} H^{--}
\to \overline{\ell}\, \overline{\ell}\, \ell\, \tau$,
where $10.8\,\%$ of $\tau$ decays into $\pi^-\nu$.

 Next,
we consider the case where
$H^{--} \to \ell\tau$ has too small branching ratio to be reliable.
 Then,
we utilize
$pp \to H^{++} H^{--}
\to \overline{\ell}\, \overline{\ell}\, \tau\, \tau$.
 As in the first case,
a pair of the same-signed $\ell$ with $M_{\ell\ell} \simeq m_{H^{\pm\pm}}^{}$
helps to reduce background events.
 For $H^{--} \to \tau\tau \to \pi^-\pi^- \nu\nu$,
the distributions%
\footnote{
 Distributions for
$H^{--} \to \tau \tau
 \to \ell_\tau^-\pi^-\, \overline{\nu}\, \nu\, \nu$
and
$H^{--} \to \tau\tau
 \to \ell_\tau^-\ell_\tau^-\, \overline{\nu}\, \overline{\nu}\, \nu\, \nu$
do not have a large difference between decays of $H^{--}_L$ and $H^{--}_R$.
 The $\ell_\tau$ denotes the $\ell$ produced via the $\tau$ decay.
}
in the collinear limit
with respect to a product $z \equiv z_1^{} z_2^{}$
of pion energy fractions%
~($z_1^{} \equiv E_{\pi_1^{}}/E_{\tau_1^{}}$
and $z_2^{} \equiv E_{\pi_2^{}}/E_{\tau_2^{}}$)
are given by
\begin{eqnarray}
&&
{\mathcal D}_{LL}^{\pi\pi}(z)
=
 \int_z^1\!\frac{dz_1^{}}{z_1^{}}\, F_L^\pi(z_1^{})\, F_L^\pi(z/z_1^{})
=
 4 \left[
    (1+z) \log\frac{1}{\,z\,} + 2z - 2
   \right] ,\\
%
%
&&
{\mathcal D}_{LL}^{\pi\pi}(z)
=
 \int_z^1\!\frac{dz_1^{}}{z_1^{}}\, F_R^\pi(z_1^{})\, F_R^\pi(z/z_1^{})
=
 4 z \log\frac{1}{\,z\,} ,
\end{eqnarray}
where ${\mathcal D}_{LL}^{\pi\pi}(z)$~(${\mathcal D}_{RR}^{\pi\pi}(z)$)
is for $H^{--}_L$~($H^{--}_R$).
 We see that the $z$ can be obtained by
$z = M_{\pi\pi}^2/m_{H^{\pm\pm}}^2$,
where $M_{\pi\pi}$ is the invariant mass
of a pair of same-signed charged pions.
 Note that
${\mathcal D}_{LL}^{\pi\pi}(z) = {\mathcal D}_{RR}^{\pi\pi}(z)$
at $z \simeq 0.2$.
 Figure~\ref{Fig:figures}~(middle)
shows the result of our simulation for this decay process.
 We see that
there is a clear difference between distributions
for $H^{--}_L$~(red dashed line) and $H^{--}_R$~(blue solid line)
even in this process.

 Finally,
we discuss about the case
where the branching ratio for $H^{--} \to \ell\ell$ is tiny.
 The fruitful process in this case is
$pp \to H^{++} H^{--}
\to \overline{\ell}\, \overline{\tau}\, \ell\, \tau$.
 The momentum of two tau leptons
can be reconstructed in the collinear limit for their decays%
~\cite{Hektor:2007uu}.
 The $H^{--}$ can be discovered
by using the invariant mass $M_{\ell\tau}$ of $\ell$ and $\tau$,
which has a peak at $m_{H^{\pm\pm}}^{}$.
 We require that a tau lepton decays into $\ell$
for the background reduction
and the other tau lepton decays into $\pi$
for the determination of the $\tau$ polarization.
 As in the first case,
the pion energy fraction $z \equiv E_\pi/E_\tau$
can be given by $z = M_{\ell\pi}^2/m_{H^{\pm\pm}}^2$.
 The generated number of $pp \to H^{++} H^{--}
\to \overline{\ell}\, \overline{\tau}\, \ell\, \tau$
is again $3\times 10^4$,
but the $\tau$ decay here is not restricted to the hadronic ones.
 The distribution of charged pions
is shown in Fig.~\ref{Fig:figures}~(right).
 It would be possible
to distinguish between $H^{--}_L$~(red dashed line)
and $H^{--}_R$~(blue solid line)
in this case also.

 In conclusion,
we can determine the chiral structure of
the Yukawa interaction of $H^{--}$
by using distributions of daughter charged pions of tau leptons
produced by the $H^{--}$ decays
if pair-produced $H^{--}$
can sufficiently decay as
$H^{++} H^{--} \to \overline{\ell}\, \overline{\ell}\, \ell\, \tau$
or $\to \overline{\ell}\, \overline{\ell}\, \tau\, \tau$
or $\to \overline{\ell}\, \overline{\tau}\, \ell\, \tau$.
 The information on the Yukawa interaction
will help to discriminate new physics models
in which neutrino masses are generated.

\section*{Acknowledgements}
 The work of H.S.\ was supported in part
by JSPS KAKENHI Grant Number~23740210.



{\small
\begin{thebibliography}{99}

\bibitem{Sugiyama:2012yw} 
  H.~Sugiyama, K.~Tsumura and H.~Yokoya,
  Phys.\ Lett.\ B {\bf 717}, 229 (2012).




\bibitem{Ref:HTM}
  W.~Konetschny and W.~Kummer,
  Phys.\ Lett.\ B {\bf 70}, 433 (1977);
%
%
  M.~Magg and C.~Wetterich,
  Phys.\ Lett.\ B {\bf 94}, 61 (1980);
%
%
  T.~P.~Cheng and L.~-F.~Li,
  Phys.\ Rev.\ D {\bf 22}, 2860 (1980);
%
%
  J.~Schechter and J.~W.~F.~Valle,
  Phys.\ Rev.\ D {\bf 22}, 2227 (1980).



\bibitem{Ref:ZBM}
  A.~Zee,
  Nucl.\ Phys.\ B {\bf 264}, 99 (1986);
%
%
  K.~S.~Babu,
  Phys.\ Lett.\ B {\bf 203}, 132 (1988).



\bibitem{Ref:tau-pol}
  K.~Hagiwara, A.~D.~Martin and D.~Zeppenfeld,
  Phys.\ Lett.\ B {\bf 235}, 198 (1990);
%
%
  A.~Rouge,
  Z.\ Phys.\ C {\bf 48}, 75 (1990).



\bibitem{Ref:tau-pol-Higgs}
  B.~K.~Bullock, K.~Hagiwara and A.~D.~Martin,
  Phys.\ Rev.\ Lett.\  {\bf 67}, 3055 (1991);
%
%
  Phys.\ Lett.\ B {\bf 273}, 501 (1991);
%
%
  Nucl.\ Phys.\ B {\bf 395}, 499 (1993).




\bibitem{Ref:tau-pol-SUSY}
  M.~M.~Nojiri,
  Phys.\ Rev.\ D {\bf 51}, 6281 (1995);
%
%
  M.~M.~Nojiri, K.~Fujii and T.~Tsukamoto,
  Phys.\ Rev.\ D {\bf 54}, 6756 (1996).



\bibitem{Hektor:2007uu} 
  A.~Hektor, M.~Kadastik, M.~Muntel, M.~Raidal and L.~Rebane,
  Nucl.\ Phys.\ B {\bf 787}, 198 (2007).
\end{thebibliography}
}
\end{document}